\documentclass{easychair}

\usepackage{doc}

\usepackage{dirtytalk}
\usepackage{color}

\title{On Quantifying and Understanding the Role of Ethics in AI Research:\\ A Historical Account of Flagship Conferences and Journals}

\author{
Marcelo Prates
\and
    Pedro Avelar
\and
   Luis C. Lamb
}

\institute{
  Federal University of Rio Grande do Sul,
  Porto Alegre, Brazil\\
  \email{morprates@inf.ufrgs.br} \\ \email{pedro.avelar@inf.ufrgs.br} \\ \email{lamb@inf.ufrgs.br}  }

\authorrunning{Prates, Avelar and Lamb}

\titlerunning{On Quantifying and Understanding the Role of Ethics in AI Research}

\begin{document}

\maketitle

\begin{abstract}
Recent developments in AI, Machine Learning and Robotics have raised concerns about the ethical consequences of both academic and industrial AI research. Leading academics, businessmen and politicians have voiced an increasing number of questions about the consequences of AI not only over people, but also on the large-scale consequences on the the future of work and employment, its social consequences and the sustainability of the planet.
In this work, we analyse the use and the occurrence of ethics-related research in leading AI, machine learning and robotics venues. In order to do so we perform long term, historical corpus-based analyses on a large number of flagship conferences and journals. Our experiments identify the prominence of ethics-related terms in published papers and presents several statistics on related topics. Finally, this research provides quantitative evidence on the pressing ethical concerns of the AI community.
\end{abstract}

% The table of contents below is added for your convenience. Please do not use
% the table of contents if you are preparing your paper for publication in the
% EPiC Series or Kalpa Publications series

%\setcounter{tocdepth}{2}
%{\small
%\tableofcontents}

%\section{To mention}
%
%Processing in EasyChair - number of pages.
%
%Examples of how EasyChair processes papers. Caveats (replacement of EC
%class, errors).

%------------------------------------------------------------------------------
\section{Introduction}

The mere notion of a universal computing mechanism raises philosophical inquires about the ultimate feasibility of building machines of human-level intelligence \cite{bostrom-ethics,ethics-society-book,russellnorvig,nature-ethics}. One of the fathers of computability theory and the first to formalize the idea of universal computation \cite{turing1937computable}, Turing began to ponder soon after his seminal paper about what means for a machine to be intelligent. His efforts culminated in the now famous Turing test \cite{turing1950computing} and the rise of the field of artificial intelligence. Concerns about the ethics and morality of computing machinery followed not long after, although also initially limited to the realm of science fiction. Acclaimed writer Isaac Asimov famously proposed his Three Laws of Robotics around the same period \cite{asimov1942runaround}, on an effort to encode norms into artificial intelligence in such a way as to prevent the rise of malicious or adversarial machines and, even then, the generally black-box nature of how the norms were encoded into the machine's brain was used to imply that it could generate unpredictable behaviours. 

Following an initial period of optimism about the future of artificial intelligence when leading scientists, including  Simon and Minsky \cite{simon1965long,minsky1967computation} predicted that Artificial General Intelligence (AGI) would be possible within the timespan of a generation, the field was struck by a wave of (sometimes intense) pessimism which lasted for decades and was later known as the \emph{AI winter} \cite{cervier1993ai}. During that period, ethical concerns about AI subsided inside the CS community, becoming more restricted to the worlds of science fiction writers, philosophers and social scientists. However, impressive machine learning results since the early 2010s are possibly turning this picture upside down faster than the computing community and the general public can cope with such changes, as pointed out by groups of experts from several leading AI countries \cite{IEEE-concerns,bostrom-chapter,royal2017machine}. 

In the timespan of half a decade, the world has seen machine learning applications progressively spread their roots into most aspects of our daily life, with smartphone intelligent personal assistants \cite{strayer2017smartphone}, targeted advertising in social networks \cite{tucker2014social} face recognition software \cite{angwin2016machine} and self-driving cars \cite{goodall2016can}. This growing phenomenon potentially raises concerns about the possibility of securing our freedom and our privacy in the face of such an interconnected and intelligent ecosystem \cite{bostrom-chapter}, as well as at which extent we can actually trust the many algorithms in the command of our daily relationship with technology not to manipulate us into making targeted decisions. 

Another pressing concern is the future of automation: will intelligent machines replace humans in the same way that automated machines took the jobs of craft workers following the industrial revolution? Moshe Vardi suggests the troubling observation that while automation is certainly eliminating traditional jobs, there is no evidence that emerging technologies create enough new jobs to compensate for those losses \cite{vardi2015humans}.  Famous technology entrepreneur Elon Musk has defended the notion of universal basic income as a possible solution  for the difficulty in distributing the wealth produced by intelligent machines, a point raised by influential businessmen; Musk has also claimed that AI poses an ``existential threat to humanity" \cite{etizioni-nytimes,muskautomation2017}. However, calls for regulating AI are ofttimes motivated by the confusion between the implications of AI science and the hypotheses raised in science fiction, as explained, for instance in \cite{etizioni-nytimes}. The IEEE Global Initiative for Ethical Considerations in Artificial Intelligence and Autonomous Systems has identified four general principles that should ``eventually serve to underpin and scaffold future norms and standards within a new framework of ethical governance'': 1) human benefit (AI should not infringe human rights) 2) responsibility (AI should be held accountable for its actions) 3) transparency and 4) education and awareness (citizens should be educated to mitigate the misuse of available AI technologies) \cite{chatila2017ieee}.

As daily life faces increasing entanglement with  information technology, it is up to AI researchers to provide safety guarantees to an increasingly anxious public. This paper contributes to both quantify and understand to which extent AI research has responded to such ethical concerns over the last decades. In particular, we are interested in how the voicing of ethical concerns by the AI community has evolved over time, and how well this process reflects the evolving demands of our society. 

The remainder of the paper is structured as follows. Next, we briefly introduce the main ethical concerns that have resulted from recent debates on ethics in AI. The topics raised in these works serve as basis for our analyses. We then describe our methodological research steps and analyse the results. Finally, we conclude and suggest further research directions.

\section{Background and Related Work} 
There are a number of ethical concerns and resulting challenges of immediate relevance faced by AI researchers \cite{nature-ethics}. For instance, face recognition software has been on the rise in the last years, and is nowadays used from everything from organizing your digital photobook \cite{helft2007google} to predicting criminal suspects \cite{klontz2013case}. The ethical validity of these technologies was brought in question  by the recent discovery of the embarrassing phenomenon of \emph{machine bias}: the process by which personal preconceptions of AI engineers can leak into projects in which they are involved. This delicate situation is perhaps best illustrated by instances of algorithmic racial bias such as Google Photos classifying dark-skinned people as gorillas \cite{garcia2016racist} or intelligent programs suggested to be negatively biased against black prisoners \cite{angwin2016machine}. 
Google's successful DeepMind team \cite{Nature-AlphaGo} has shown that machine learning based systems can achieve superhuman performance in the challenging domain of game playing, in which algorithms were trained by `supervised learning from human experts' and `reinforcement learning from self-play'. 

Rossi has pointed out that humans and machines will have to reach common agreement on collective decisions, either by consensus or negotiated compromises when acting in a common environment \cite{rossi2015safety}. Researchers have also revived debates concerning the controversial field of physiognomy, with many people asking whether artificial intelligence even \emph{should} try and classify people's sexual orientation according to their facial features \cite{wang2017deep}.
Among the many challenges identified in \cite{royal2017machine} better transparency, interpretability and explainability of AI technologies \cite{bostrom-ethics,doran2017does} would lead to improved acceptance of AI technologies in society. In addition, in order to increase public confidence in AI, algorithms and systems must be made accountable; AI professionals are already seen to a certain extent as as responsible for their (desirably explainable) actions.\footnote{When asked about  \emph{``who should be held accountable when machine learning `goes wrong'''}, 32\% of the public attributed such responsibility to \emph{``the organisation the operator and machine work for''}. \cite{royal2017machine}, p. 96.}

As the prominence of artificial intelligence and particularly machine learning systems in our society rapidly increases, a large number of ethical concerns become pressing. Addressing these issues is a problem in itself, as the public awareness about the nature and operation of machine learning systems seems to be fairly limited. When inquired about the topic, as few as 9\% of the participants declared having heard the topic \say{machine learning} and only 3\% said they knew a great deal or fair amount about the field. By contrast, 76\% had heard of computers that can recognize speech and answer questions and 89\% had heard of at least one of the eight examples of machine learning used in the survey \cite{royal2017machine}. This possibly suggests that people are generally familiar with the \emph{applications} of machine learning (ML)  while ignoring the fundamental principles behind them.

\section{Ethical Concerns Impacting Artificial Intelligence and Machine Learning}
One of the oldest and most prominent concerns impacting automation is the replacement of human workforce by intelligent systems. This is a delicate topic, with people tending to disagree about where to draw the line concerning the adoption of robots in the workplace. On the one hand, people are content with robots replacing human workers in positions which could be considered harmful or dangerous, but at the same time the use of robots in personal or caring roles is viewed disfavorably due to the fear of losing human-to-human contact \cite{castell2014public}. On a study conducted by the Royal Society, public opinions about automation by machine learning systems were also mixed  \cite{royal2017machine}. On the positive side, people think that machine learning systems could be more objective than human users, helping to avoid cases of human error which arise when decision-makers are tired or emotionally vulnerable. They also believe that machine learning systems could be more accurate than human professionals, for example in conducting medical diagnoses. The perspective of automation bringing efficiency to the public sector is viewed favorably, as well as its potential to catalyze economic growth and tackle large scale societal challenges such as climate change. Nevertheless, people fear that machine learning can lead to physical harm to human beings, for example in accidents involving autonomous vehicles. The replacement of humans by machines in the workplace inspires fear of unemployment as well as of our over-reliance on them to make diagnoses. The issue of human replacement was raised spontaneously and frequently over the course of the study, suggesting that it is a sensitive matter for the public.

The employment of ML in the automation of key services raises concerns about the effects of depersonalization and consumer misdirection. People feel that, lacking qualities such as human empathy and personal engagement, ML systems could have an effect on the depersonalization in the delivery of key services. There is the fear that ML-powered targeted ads could mislabel or inadvertently stereotype consumers, and that the prominence of ML in the Internet could create an algorithmic bubble which would filter challenging opinions, experiences or interactions \cite{royal2017machine}.

Privacy is a sensitive and controversial topic, with people's levels of concern about data privacy generally varying according to the circumstances \cite{royal2017machine}. The issue is further complicated by the potential of ML to uncover sensitive relationships with limited data, as suggested by a PNAS study showing that a list of attributes including sexual orientation, ethnicity, religion, political views, intelligence and gender can be inferred from publicly accessible digital records such as Facebook likes \cite{kosinski2013private}. The take-home lesson is that even if sensitive attributes are explicitly removed from the training data, the remaining attributes can still link to them.

A recent concern is that of \emph{machine bias}, which has received increasingly more attention as trained statistical models rapidly become the default in various applications. A number of studies has suggested that ML can fall victim to the same prejudices, stereotypes or biases possessed by their creators/programmers, with implications to racism and sexism in our society \cite{garcia2016racist,angwin2016machine}. Intelligent systems which become negatively biased against minorities because of ill-designed training sets are bad enough, but we should also consider that \emph{even when machine learning uncovers a valid association, its use in recommendation systems may be controversial}.

In the age of autonomous vehicles, one of the most pressing concerns becomes that of \emph{accountability}. If a self-driving car is involved in an accident, who should bear the blame? In a more general sense, \emph{who should be accountable when machine learning systems goes wrong}? Many AI models effectively become black boxes upon training, and their methods and functioning become difficult to interpret -- because the underlying algorithms of ML systems learn from training data, simply knowing the underlying program is different from knowing which features it will weight on the most. It is somewhat accepted that ML systems should be judged by their accuracy, and that ML systems which are more accurate than their human counterparts should be considered for replacement. But it could also be argued that if the decisions and predictions at hand have a significant impact, then understanding how they were computed is possibly more important than higher levels of accuracy.

\section{Methodology}\label{sec:methodology}

To achieve a measure of how much Ethics in AI is discussed we carried out extensive analyses of  the mainstream AI venues. In our experiments, we search for ethics-related terms in the titles of papers in flagship AI, machine learning and robotics conferences and journals. 
The terms we search for were based on the issues exposed and identified in \cite{bossmann2016top,burton2017ethical,royal2017machine}, 
and also on the topics called for discussion in the First AAAI/ACM Conference on AI, Ethics, and Society. The \textbf{ethics keywords} used were the following: \textit{Accountability, Accountable, Employment, Ethic, Ethical, Ethics, Fool, Fooled, Fooling, Humane, Humanity, Law, Machine bias, Moral, Morality, Privacy, Racism, Racist, Responsibility, Rights, Secure, Security, Sentience, Sentient, Society, Sustainability, Unemployment and Workforce}.

The list was larger, however, during a first analysis of the data we found out that some of the keywords that were to be used provided too many articles in which these words were used in ways unrelated to ethics in AI research. Some examples are the keywords control and controllable in Robotics venues: Since their use is generally attributed to the context of \emph{control systems}, they should not be considered in the analyses and the keyword social, which generally was used as a part of ``social networks''.
After the identification/discovery of these keywords we filtered the results further by manually removing papers with keyword matches whose context was not ethics-related. 

If we want to assess the level of attention or relevance given to ethical issues by the AI research community, it is necessary to have some form of baseline. With this in mind, we proposed two additional keyword sets encompassing classical AI terms such as reasoning, planning, learning, etc as well as trending topics such as convolution neural networks, deep learning, SLAM, etc. By comparing the evolution of the frequencies in which keywords from these three different categories (ethics, classical, trending) match paper titles, one can gain insights about what the AI and robotics research communities have prioritized over time.

The classical and trending keyword sets were compiled from the areas in the most cited book on AI by Russell and Norvig \cite{russellnorvig} and from curating terms from the keywords that appeared most frequently in paper titles over time in the venues. The keywords chosen for the \textbf{classical keywords} category were: \textit{Cognition, Cognitive, Constraint satisfaction, Game theoretic, Game theory, Heuristic search, Knowledge representation, Learning, Logic, Logical, Multiagent, Natural language, Optimization, Perception, Planning, Problem solving, Reasoning, Robot, Robotics, Robots, Scheduling, Uncertainty and Vision}. The curated \textbf{trending keywords} were: \textit{Autonomous, Boltzmann machine, Convolutional networks, Deep learning, Deep networks, Long short term memory, Machine learning, Mapping, Navigation, Neural, Neural network, Reinforcement learning, Representation learning, Robotics, Self driving, Self-driving, Sensing, Slam, Supervised/Unsupervised learning and Unmanned}.\footnote{All datasets from the paper's experiments  will be made available in the final version. We omit any links to the data to prevent author identification.}

Since abstracts in text form were available for a smaller number of papers, as a way of validating that our results would remain true in the case that the corpora analysis was made wholly on abstracts, we observed the conditional probability that a word would appear on a title, given that it appears on a abstract on those papers that had textual abstracts available. This was done filtering stopwords, and was done for the set of keywords that are not  ethics-related and for those that are -- we call the first $P(T|A)_{K}$ and $P(T|A)_{E}$. After running this we observed an $P(T|A)_{E}$ bigger than $P(T|A)_{K}$, with $P(T|A)_{E} = 11.53\%$ and $P(T|A)_{K} = 8.71\%$. The way this is put, one can say that if we count the occurrences only in titles, we can expect to under-sample ethics less than we under-sample the rest of the keywords; thus if we identify a gap where ethics keywords appear less in titles, this gap would be only intensified if we expanded to abstracts.
A simple way to visualise this is that given a number of measured occurrences of ethics related keywords $\#E_{m}$ and non-ethics related keywords $\#K_{m}$ we can expect their true values $\#E_{t}$ and $\#K_{t}$ be in a relation like:
\[\#K_{m} \approx  P(T|A)_{k} * \#K_{t}\]
\[\#E_{m} \approx  P(T|A)_{e} * \#E_{t}\]
Thus, if $P(T|A)_{e} > P(T|A)_{k}$ one can expect that $\#K_{m}/\#E_{m} > \#K_{t}/\#E_{t}$ -- that is, the proportion of non-ethics related keywords would only increase if all abstracts were considered and the probabilities stayed the same.

\section{Experimental Analyses and Results}

The following statistics were computed on a dataset of a total of $110,108$ papers, encompassing $59,352$ conference and $50,756$ journal entries (see Table \ref{tab:samplesize}).
The experiments and results summarized here are  stratified into three groups:\\
(1) The \emph{AI} group contains papers from the main Artificial Intelligence and Machine Learning conferences such as AAAI, IJCAI, ICML, NIPS and also from both the Artificial Intelligence Journal and the Journal of Artificial Intelligence Research (JAIR). \\
(2) The \emph{Robotics} group contain papers published in the IEEE Transactions on Robotics and Automation (now known as IEEE Transactions on Robotics), ICRA and IROS. \\
(3) The \emph{CS} group contains papers published in the mainstream Computer Science venues such as the Communications of the ACM, IEEE Computer, ACM Computing Surveys and the ACM and IEEE Transactions.
\begin{table}[h]
    \centering
    \label{multiprogram}
    \begin{tabular}{|p{1cm}|p{1cm}|p{1cm}|p{1cm}|p{1cm}|p{1cm}|}
        \hline
        \multicolumn{6}{|c|}{Conferences} \\ \hline 
        AAAI & IJCAI & NIPS & ICML & ICRA & IROS \\ \hline
        $7,179$ & $7,723$ & $6,509$ & $3,568$ & $19,368$ & $15,005$ \\ \hline
        \multicolumn{6}{|c|}{Journals} \\ \hline
        ACM Trans. & Comm. ACM & IEEE. Computer & JAIR & IEEE Trans. AI & Artif. Intell. \\ \hline
        $18,199$ & $11,394$ & $6,694$ & $972$ & $10,731$ & $2,766$ \\ \hline
    \end{tabular}
    \caption{Sample sizes in number of papers for the analysed venues.}
    \label{tab:samplesize}
\end{table}
For brevity, a number of similar venues were grouped into a single category. In Table \ref{tab:samplesize}, the column 
``IEEE Trans. AI'' groups together a number of AI-related IEEE Transactions. They are: IEEE Trans. on Affective Computing, IEEE Trans. on Audio, Speech and Language Processing, IEEE Trans. on Cognitive and Developmental Systems, IEEE Trans. on Computational Intelligence and AI in Games, IEEE Trans. on Emerging Topics in Computing, IEEE Trans. on Fuzzy Systems, IEEE Trans. on Intelligent Systems, IEEE Trans. on Neural Networks and Learning Systems.           

\begin{figure}[h]
\centering
\includegraphics[width=0.75\linewidth]{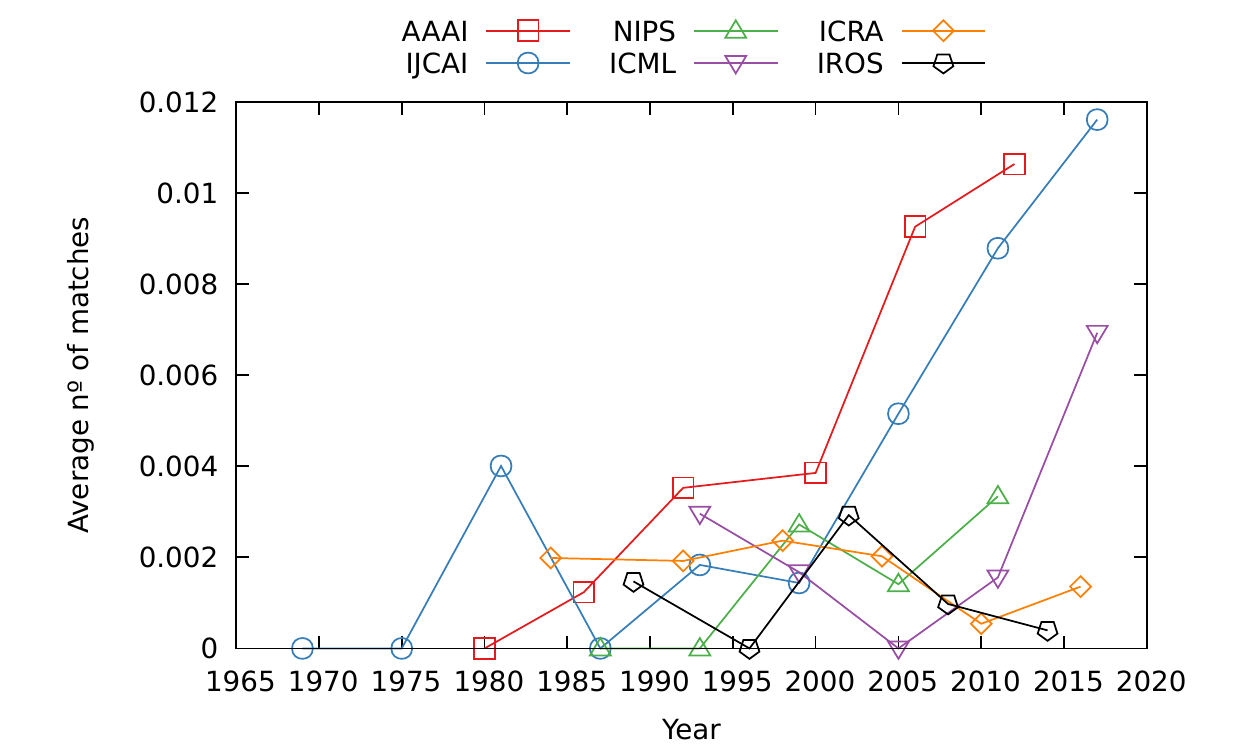}
\caption{Frequency of the selected ethics-related keywords (see Sec.~\ref{sec:methodology} for the list) per five year interval in paper titles for five of the leading AI (AAAI, IJCAI, NIPS and ICML) and Robotics (ICRA and IROS) conferences.}
\label{fig:titles-confs}
\end{figure}

For each publication, we compute the number of times each of our selected keywords occurs in its title. These statistics are grouped first by venue and afterwards by year of publication (or, in some cases, publications are grouped by five year intervals). From the statistics for each keyword we also compute the total number of matches, which is averaged over all samples. For example, the y-axis of Figure \ref{fig:titles-confs} corresponds to the average number of keyword matches throughout all publications of the same venue per five year interval.

\begin{figure}[htb]
\centering
\includegraphics[width=0.75\linewidth]{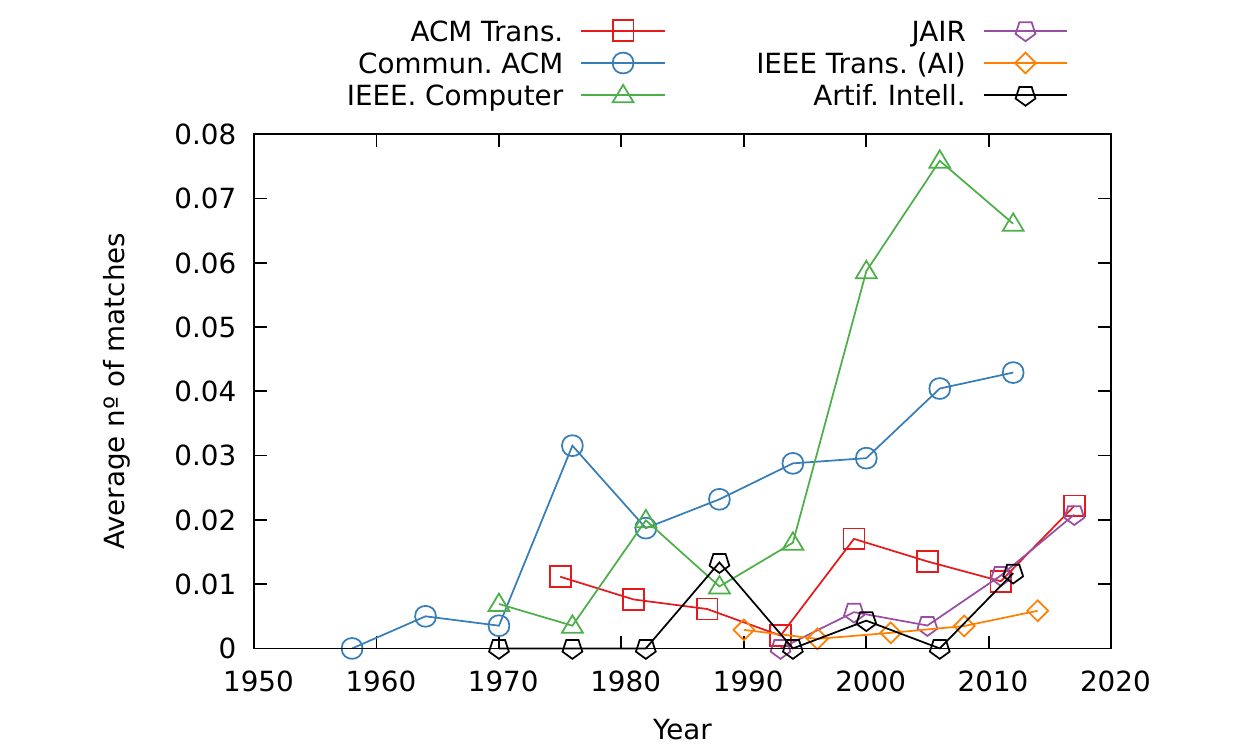}
\caption{Frequency of the selected ethics-related keywords (see Sec.~\ref{sec:methodology} for the list) per five year interval in paper titles for leading computing journals.}
\label{fig:titles-journals}
\end{figure}

Figure \ref{fig:titles-confs} shows the evolution of keyword frequencies for some of the leading AI and Robotics conferences. While the trend for AAAI and IJCAI suggests a growing interest for ethics related themes by part of the AI community, the data for NIPS, ICML, ICRA and IROS is not conclusive. The scale of keyword frequencies, ranging up to $0.012$ further suggests that ethical concerns receive little attention by these venues. Computing journals seem to devote more attention to these issues, with up to $0.08$ of paper titles matches with ethics-related keywords as Figure \ref{fig:titles-journals} shows. 

\begin{figure}[h]
\centering
\includegraphics[width=0.75\linewidth]{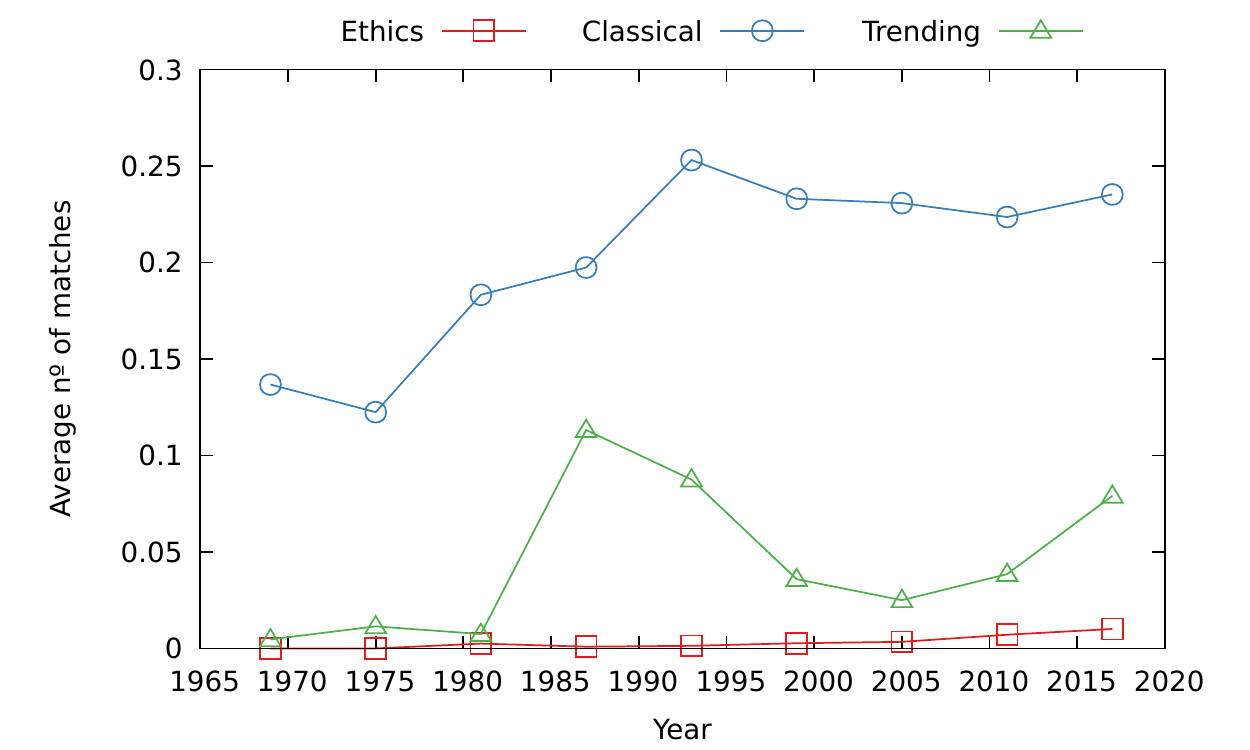}
\caption{Comparison of the frequencies of ethics-related keywords with classical and trending AI keywords (see Sec.~\ref{sec:methodology} for the lists) per five year interval in paper titles for leading AI conferences (AAAI, IJCAI, NIPS, ICML).}
\label{fig:ai-ethics-classical-trending}
\end{figure}

\begin{figure}[h]
\centering
\includegraphics[width=0.75\linewidth]{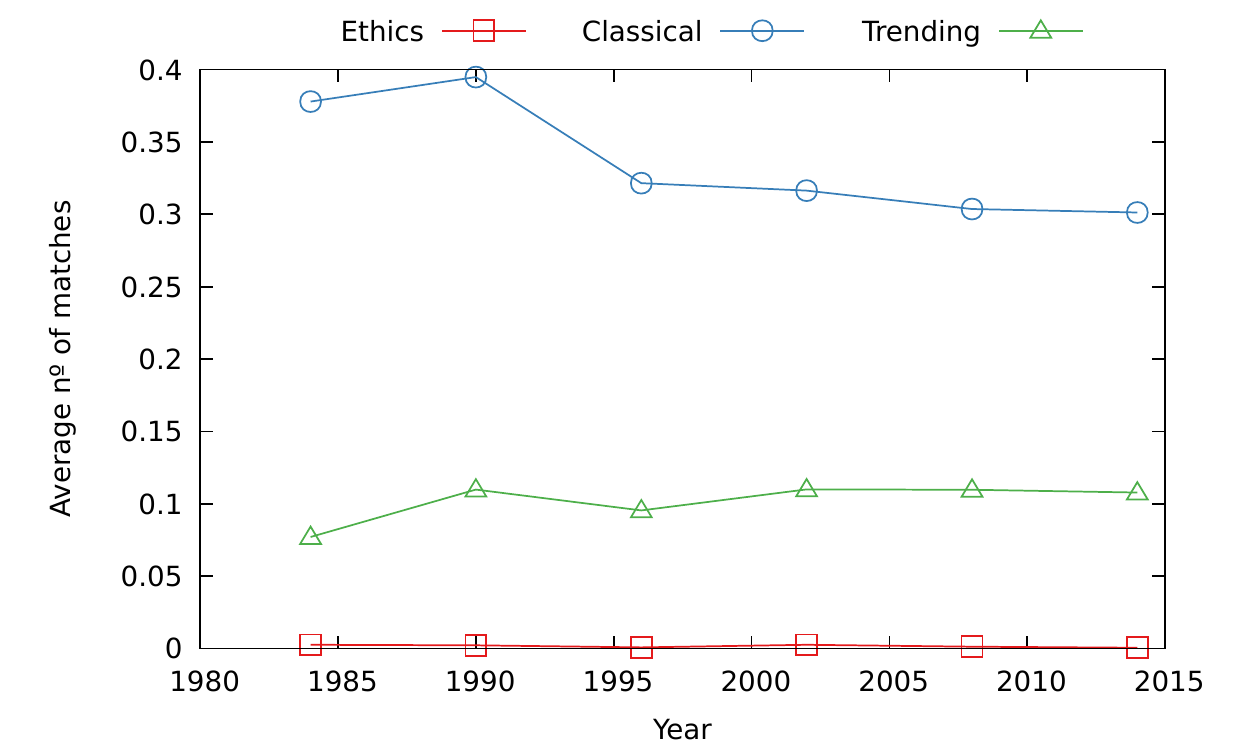}
\caption{Comparison of the frequencies of ethics-related keywords with classical and trending AI keywords (see Sec.~\ref{sec:methodology} for the lists) per five year interval in paper titles for leading Robotics conferences (ICRA, IROS).}
\label{fig:robotics-ethics-classical-trending}
\end{figure}

When ethics-related keyword frequencies are compared with those of classical or trending AI terms, we get a possibly troubling picture. The supremacy of consecrated computing topics in these venues is to be expected, but Figure \ref{fig:ai-ethics-classical-trending} shows the extent to which popular technologies such as deep learning, Boltzmann machines, convolutional networks, self driving cars, etc. overshadow the ethical concerns expressed on paper titles of the top AI conferences. The peak in the \emph{trending} curve in the late 80s is explained by the neural network developments at that time, and one can see that the same terms are on the rise once again since the early 2010s -- although unfortunately this is not accompanied by a substantial increase in ethical concerns. The data for robotics conferences shown in Figure \ref{fig:robotics-ethics-classical-trending} suggests an even larger gap between ethics-related topics and trending technologies.

\begin{figure}[h]
\centering
\includegraphics[width=0.75\linewidth]{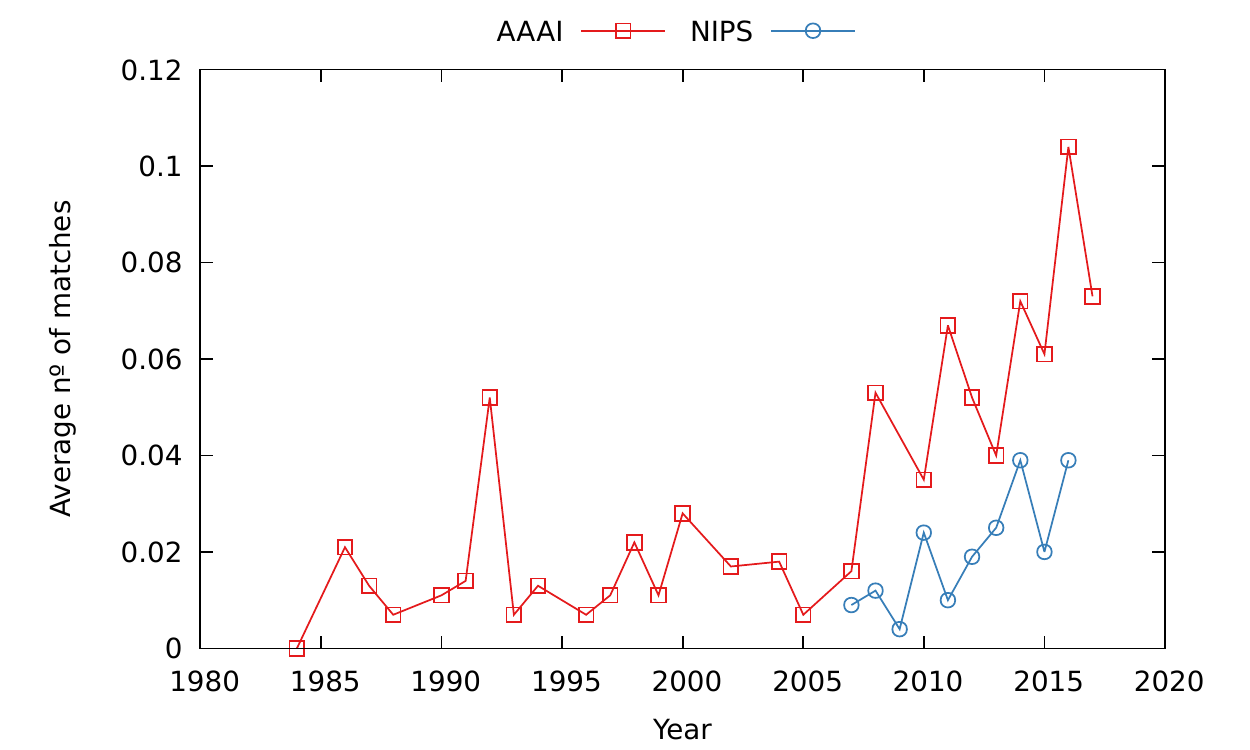}
\caption{Frequency of the selected ethics-related keywords (see Sec.~\ref{sec:methodology} for the list) per year in AAAI and NIPS paper abstracts ranging from 1984 to 2017.}
\label{fig:abstracts-aaai-nips}
\end{figure}

For AAAI and NIPS we were able to collect statistics about keyword frequencies in paper abstracts as well as their titles. Figure \ref{fig:abstracts-aaai-nips} compares the evolution in the frequency of ethics-related keyword matches for both conferences, once again suggesting that perhaps too little attention is devoted to these topics by two of the leading AI venues. Incorporating abstracts into our corpora yields almost no noticeable differences in match frequencies, with AAAI and NIPS frequencies peaking close to $0.01$ and $0.004$ respectively towards the end of the current decade.

\begin{figure}[h]
\centering
\includegraphics[width=0.75\linewidth]{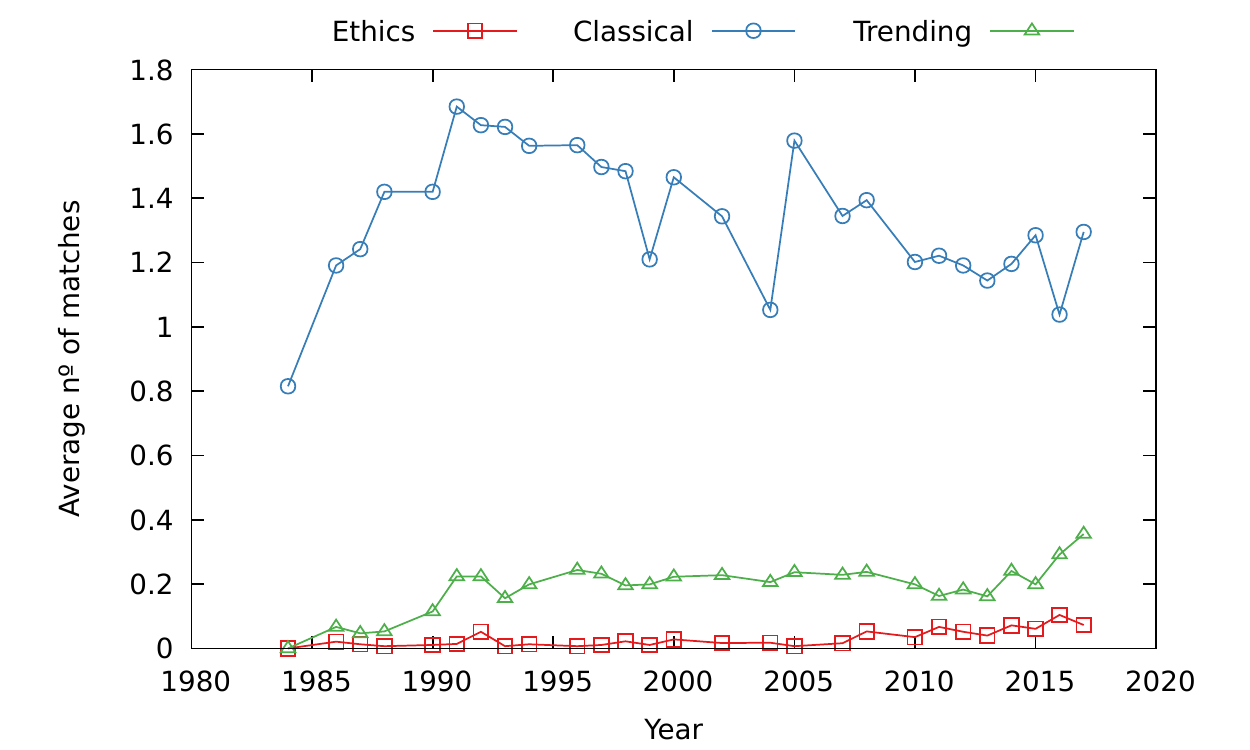}
\caption{Frequency of the selected ethics-related keywords (see Sec.~\ref{sec:methodology} for the list) per year in AAAI paper abstracts ranging from 1984 to 2017.}
\label{fig:aaai-abstracts-ethics-classical-trending}
\end{figure}

\begin{figure}[h]
\centering
\includegraphics[width=0.75\linewidth]{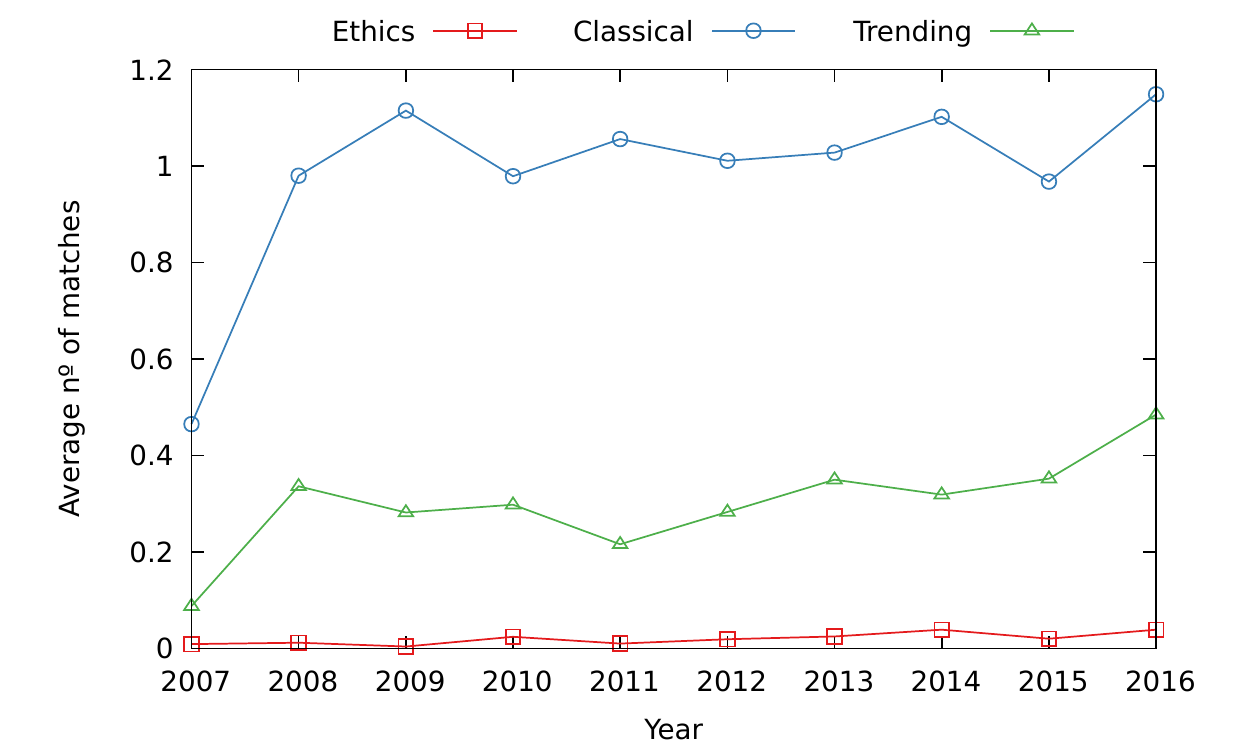}
\caption{Frequency of the selected ethics-related keywords (see Sec.~\ref{sec:methodology} for the list) per year in NIPS paper abstracts ranging from 2007 to 2017.}
\label{fig:nips-abstracts-ethics-classical-trending}
\end{figure}

Figures \ref{fig:aaai-abstracts-ethics-classical-trending} and \ref{fig:nips-abstracts-ethics-classical-trending} further show how the voicing of ethical concerns compares with the frequency of consecrated CS terms and trending/emerging technologies for AAAI and NIPS respectively, repeating the overshadowing of ethics-related discussions by popular topics observed in Figures \ref{fig:ai-ethics-classical-trending} and \ref{fig:robotics-ethics-classical-trending}. Tables \ref{tableresults_conferences_stats} and \ref{tab:results_journals_stats} illustrate a more complete picture of the data collected and analyzed in this paper. Notice that some years in these tables have been removed due to the absence of keywords matches or papers in these years.

%%%%%%%%%%%%%%%%%%%%%%%%%%%%%%%
%table conferences here
\begin{table}[h]
\scriptsize
\centering
\begin{tabular}{|c|c|c|c|c|c|c|}
\hline
Year & AAAI & IJCAI & NIPS & ICML & ICRA & IROS \\ \hline
%81 & -         & $0 / 0$     & -       & -       & -       & - \\ \hline
%82 & $0 / 0$     & -         & -       & -       & -       & - \\ \hline
83 & $0 / 0$    & $.004 / 1$    & -       & -       & -       & - \\ \hline
%84 & $0 / 0$     & -         & -       & -       & $0 / 0$   & - \\ \hline
85 & -        & $.008 / 2$    & -       & -       & $0 / 0$   & - \\ \hline
%86 & $0 / 0$     & -         & -       & -       & $0 / 0$   & - \\ \hline
87 & $0 / 0$    & $0 / 0$     & $0 / 0$     & -       & $.003/ 1$   & - \\ \hline
88 & $.007 / 1$   & -         & $0 / 0$     & -       & $.003/ 1$   & - \\ \hline
89 & -        & $0 / 0$     & $0 / 0$     & -       & $.004/ 1$   & $.011 / 1$ \\ \hline
%90 & $0 / 0$     & -         & $0 / 0$     & -       & -       & $0 / 0$ \\ \hline
%91 & $0 / 0$     & $0 / 0$       & $0 / 0$     & -       & -       & $0 / 0$ \\ \hline
92 & $.007 / 1$   & -         & $0 / 0$     & -       & $.007/ 3$   & $.004 / 1$ \\ \hline
%93 & $0  / 0$    & $0 / 0$       & $0 / 0$     & $0/ 0$    & -       & $0 / 0$ \\ \hline
94 & $.003 / 1$   & -         & $0 / 0$     & $.023/ 1$   & $0 / 0$   & $0 / 0$ \\ \hline
95 & -        & $0 / 0$       & $0 / 0$     & $0 / 0$   & $.004 / 2$  & - \\ \hline
%96 & -         & -         & $0 / 0$     & $0 / 0$   & $0 / 0$   & $0 / 0$ \\ \hline
97 & -        & $.011 / 1$    & $0 / 0$     & $0 / 0$   & $0 / 0$   & $0 / 0$ \\ \hline
98 & -        & -         & $0 / 0$     & $0 / 0$   & $.002 / 1$  & $0 / 0$ \\ \hline
99 & -        & $0 / 0$       & $.007 / 1$  & $0 / 0$   & $0 / 0$   & $0 / 0$ \\ \hline
00 & $0 / 0$    & -         & $0 / 0$   & $.007/ 1$   & $.002 / 1$  & $0 / 0$ \\ \hline
01 & -        & $0 / 0$       & $0 / 0$   & $0 / 0$   & $.003 / 2$  & $0 / 0$ \\ \hline
02 & -        & -         & $.005 / 1$  & $0 / 0$   & $.001 / 1$  & $0 / 0$ \\ \hline
03 & -        & $.003 / 1$    & $.005 / 1$  & $0 / 0$   & $.006 / 4$  & $.003 / 2$ \\ \hline
04 & $1 / 2$    & -         & $0 / 0$   & $0 / 0$   & $.002 / 2$  & $.002 / 1$ \\ \hline
05 & $0 / 0$    & $.011 / 4$    & $0 / 0$   & $0 / 0$   & $0 / 0$   & $.005 / 3$ \\ \hline
06 & $0.005 / 2$  & -         & $.005 / 1$  & $0 / 0$   & $.004 / 3$  & $.003 / 3$ \\ \hline
07 & $0 / 0$    & $.004 / 2$    & $0 / 0$   & $0 / 0$   & $0 / 0$   & $.004 / 3$ \\ \hline
08 & $.017 / 6$   & -         & $0 / 0$   & $0 / 0$   & $.003 / 2$  & $.003 / 2$ \\ \hline
09 & -        & $0 / 0$       & $0 / 0$   & $0 / 0$   & $.003 / 2$  & $0 / 0$ \\ \hline
10 & $.006 / 2$   & -         & $.003 / 1$  & $0 / 0$   & $.001 / 1$  & $.002 / 2$ \\ \hline
11 & $.019 / 6$   & $.01 / 5$     & $0 / 0$   & $0 / 0$   & $.002 / 2$  & $0 / 0$ \\ \hline
12 & $.011 / 4$   & -         & $.003 / 1$  & $0 / 0$   & $0 / 0$   & $.001 / 1$ \\ \hline
13 & $.008 / 2$   & $.01 / 5$     & $.003 / 1$  & -       & $0 / 0$   & $0 / 0$ \\ \hline
14 & $.010 / 5$   & -         & $.005 / 2$  & $0 / 0$   & $0 / 0$   & $0 / 0$ \\ \hline
15 & $.004 / 3$   & $.01 / 7$     & $.002 / 1$  & $.007/ 2$   & $0 / 0$   & $0 / 0$ \\ \hline
16 & $.017 / 12$  & $.004 / 3$    & $.005 / 3$  & $0 / 0$   & $.001 / 1$  & $.001 / 1$ \\ \hline
17 & $0.01 / 8$   & $.011 / 9$    & $.004 / 3$      & $.007/ 3$   & $.001 / 1$  & $0 / 0$ \\ \hline
\end{tabular}
\caption{Average and total number of matches of the selected ethics-related keywords per year in paper titles for six leading AI (AAAI, IJCAI, NIPS, ICML) and Robotics (ICRA, IROS) conferences from $1981$ to $2017$. Omitted years had no occurrence of the keywords}
\label{tableresults_conferences_stats}
\end{table}
%%%%%%%%%%%%%%%%%%%%%%%%
%%%%table journals here
\begin{table}[h]
\scriptsize
\centering
\begin{tabular}{|c|p{0.8cm}|p{0.8cm}|p{0.8cm}|p{0.8cm}|p{0.8cm}|p{0.8cm}|}
\hline
Year & ACM Trans. & Comm. ACM & IEEE Computer & JAIR & IEEE Trans. AI & Artif. Intell. \\ \hline
81 & $.00/ 0$ & $.00/ 0$  & $.00/ 0$  & -       & -       & $.00/ 0$ \\ \hline
82 & $.00/ 0$ & $.00/ 0$  & $.02/ 2$  & -       & -       & $.00/ 0$ \\ \hline
83 & $.00/ 0$ & $.00/ 0$  & $.00/ 0$  & -       & -       & $.03/ 1$ \\ \hline
84 & $.00/ 0$ & $.00/ 0$  & $.02/ 2$  & -       & -       & $.00/ 0$ \\ \hline
85 & $.00/ 0$ & $.00/ 0$  & $.00/ 0$  & -       & -       & $.00/ 0$ \\ \hline
86 & $.00/ 0$ & $.01/ 1$  & $.00/ 0$  & -       & -       & $.00/ 0$ \\ \hline
87 & $.01/ 1$ & $.00/ 0$  & $.01/ 1$  & -       & -       & $.00/ 0$ \\ \hline
88 & $.01/ 1$ & $.00/ 0$  & $.13/ 8$  & -       & -       & $.00/ 0$ \\ \hline
89 & $.00/ 0$ & $.00/ 0$  & $.08/ 8$  & -       & -       & $.02/ 1$ \\ \hline
90 & $.00/ 0$ & $.00/ 0$  & $.00/ 0$  & -       & .$26/ 9$    & $.03/ 2$ \\ \hline
91 & $.00/ 0$ & $.01/ 1$  & $.04/ 4$  & -       & $.30/ 22$   & $.00/ 0$ \\ \hline
92 & $.00/ 0$ & $.00/ 0$  & $.02/ 2$  & -       & $.32/ 33$   & $.01/ 1$ \\ \hline
93 & $.00/ 0$ & $.00/ 0$  & $.00/ 0$  & $.00/ 0$    & $.20/ 26$   & $.04/ 5$ \\ \hline
94 & $.00/ 1$ & $.03/ 4$  & $.00/ 0$  & $.00/ 0$    & $.17/ 22$   & $.02/ 2$ \\ \hline
95 & $.00/ 0$ & $.01/ 1$  & $.00/ 0$  & $.00/ 0$    & $.14/ 30$   & $.01/ 1$ \\ \hline
96 & $.00/ 0$ & $.00/ 0$  & $.06/ 8$  & $.00/ 0$    & $.20/ 41$   & $.00/ 0$ \\ \hline
97 & $.00/ 0$ & $.00/ 1$  & $.01/ 2$  & $.05/ 1$    & $.23/ 48$   & $.03/ 3$ \\ \hline
98 & $.01/ 3$ & $.00/ 0$  & $.00/ 0$  & $.00/ 0$    & $.23/ 47$   & $.04/ 4$ \\ \hline
99 & $.01/ 3$ & $.01/ 3$  & $.02/ 3$  & $.00/ 0$    & $.17/ 40$   & $.01/ 1$ \\ \hline
00 & $.01/ 2$ & $.00/ 0$  & $.01/ 2$  & $.00/ 0$    & $.12/ 28$   & $.01/ 1$ \\ \hline
01 & $.01/ 3$ & $.00/ 1$  & $.00/ 0$  & $.04/ 1$    & $.15/ 38$   & $.02/ 2$ \\ \hline
02 & $.02/ 6$ & $.00/ 1$  & $.00/ 1$  & $.00/ 0$    & $.15/ 39$   & $.03/ 2$ \\ \hline
03 & $.00/ 1$ & $.01/ 2$  & $.00/ 1$  & $.00/ 0$    & $.12/ 29$   & $.00/ 0$ \\ \hline
04 & $.00/ 1$ & $.00/ 1$  & $.01/ 2$  & $.00/ 0$    & $.13/ 32$   & $.02/ 1$ \\ \hline
05 & $.02/ 10$  & $.00/ 0$  & $.01/ 2$  & $.00/ 0$    & $.11/ 29$   & $.03/ 2$ \\ \hline
06 & $.01/ 5$ & $.00/ 0$  & $.02/ 6$  & $.00/ 0$    & $.10/ 47$   & $.00/ 0$ \\ \hline
07 & $.01/ 9$ & $.01/ 3$  & $.01/ 2$  & $.02/ 1$    & $.14/ 76$   & $.00/ 0$ \\ \hline
08 & $.01/ 6$ & $.00/ 0$  & $.00/ 0$  & $.03/ 2$    & $.15/ 78$   & $.00/ 0$ \\ \hline
09 & $.01/ 5$ & $.00/ 1$  & $.00/ 0$  & $.02/ 1$    & $.17/ 88$   & $.02/ 1$ \\ \hline
10 & $.00/ 2$ & $.01/ 2$  & $.00/ 0$  & $.00/ 0$    & $.05/ 32$   & $.00/ 0$ \\ \hline
11 & $.00/ 1$ & $.00/ 1$  & $.01/ 1$  & $.00/ 0$    & $.12/ 79$   & $.01/ 1$ \\ \hline
12 & $.01/ 9$ & $.00/ 1$  & $.00/ 0$  & $.02/ 1$    & $.10/ 74$   & $.00/ 0$ \\ \hline
13 & $.01/ 9$ & $.01/ 4$  & $.00/ 0$  & $.00/ 0$    & $.08/ 61$   & $.00/ 0$ \\ \hline
14 & $.01/ 16$  & $.00/ 0$  & $.01/ 2$  & $.00/ 0$    & $.12/ 77$   & $.02/ 1$ \\ \hline
15 & $.03/ 34$  & $.01/ 3$  & $.01/ 1$  & $.00/ 0$    & $.13/ 107$  & $.03/ 2$ \\ \hline
16 & $.03/ 43$  & $.02/ 5$  & $.01/ 2$  & $.02/ 1$    & $.10/ 78$   & $.01/ 1$ \\ \hline
17 & $.02/ 29$  & $.02/ 4$  & $.03/ 5$  & $.02/ 1$    & $.14/ 94$   & $.02/ 2$ \\ \hline
\end{tabular}
\caption{Average and total number of matches of the selected ethics-related keywords per year in paper titles for six groups of the leading computing journals from $1981$ to $2017$. IEEE journals in AI are grouped into IEEE Trans. AI.}
\label{tab:results_journals_stats}
\end{table}
%\begin{figure}[!h]
%\centering
%\includegraphics[width=1\linewidth]{pictures/google-trends.pdf}
%\caption{Public interest over time of some of the ethics-related keywords considered in this work. Data from Google Trends.}
%\label{fig:google-trends}
%\end{figure}
\pagebreak
\section{Conclusions}
In this paper, we carried out an investigation of the long-term prominence of ethics related research in flagship AI venues. In order to do so, we performed corpora analyses on a large number of artificial intelligence, machine learning, and robotics top conferences and journals. 
The focus on ethical consequences and implications of AI has been in the field's research agenda since its dawn. However, specific interest on ethics-related research topics has not been consistent over the decades. The experiments identified a relatively low attention of the AI community with respect to ethical consequences of AI along the decades, as shown by our data analyses. 

One could argue that there have been seminars and smaller workshops on particular topics associated with Ethics in AI and related areas, which would contradict the low percentage and absolute numbers of ethics-related research papers in AI venues. 
However, our results show that over the last decades ethical issues have not been present at the \emph{main tracks} of the 
flagship AI venues. Although workshops and smaller events may raise awareness among researchers and professionals, given the relevance and prominence AI technology has achieved in society, one can argue that ethics-related research 
should have perhaps dedicated tracks alongside the technical contents in the leading AI, machine learning and robotics venues.

Even though the prospects of achieving artificial general intelligence (or strong AI) and the singularity still seem far in the horizon, the ever expanding influence of intelligent systems in our society strongly suggests that ethics should be very much a present-day concern for AI research, and perhaps more so today than in any other point in the history of the field. In addition, the development of AI systems  and tools raises several issues related to fairness, (algorithmic) accountability \cite{nature-ethics} and justice \cite{pitt}.

As clearly identified by the experts in the Royal Society report \cite{royal2017machine}, public concern about transparency, accountability and consequences of AI in general, and machine learning in particular require that both current and future researchers take into account the ethical consequences of their research.
In this context, our work has contributed to not only identify the many faces of ethics in AI research over the years, but also has shown that current and flagship AI venues and researchers still dedicate a limited amount of their research focus to ethics in AI, machine learning and robotics. 

The identification of relevant research topics, or relative lack of attention thereof, opens several opportunities and challenges for the AI community, which will contribute to the development of accountable, sustainable and ethical systems and technologies with positive impact in human life and society. The societal demand for transparency and interpretability of AI systems also require increasing awareness of the research community. We believe this research contributes toward these aims, by providing experimental evidence of the historical evolution of ethics in AI research.
\section{Acknowledgements}
This work is partly sopported by the Brazilian Research Agencies CAPES, CNPq and FAPERGS.
\label{sect:bib}
\bibliographystyle{plain}
\bibliography{easychair}

%------------------------------------------------------------------------------

%------------------------------------------------------------------------------
% Index
%\printindex

%------------------------------------------------------------------------------
\end{document}